\documentclass{PoS}

\def \et {E_{T}}

\def  \met {\not\!\!\et }

\title{Higgs searches at the Tevatron}

\ShortTitle{Higgs searches at the Tevatron}

\author{\speaker{Ben Kilminster}\thanks{on behalf of the CDF and D0 collaborations}\\
        Fermilab\\
        E-mail: \email{bjk AT fnal.gov}}

\abstract{We present combined CDF and D0 searches for the Standard Model (SM) and Minimal Supersymmetric Standard Model (MSSM) Higgs boson using up to 6.7 fb$^{-1}$ of integrated luminosity from 1.96 TeV proton-antiproton collisions at the Tevatron.  Specialized searches for Higgs bosons produced via gluon fusion, associated boson production, and vector boson fusion, decaying to $b\bar{b}$, $W^{+}W^{-}$, $\tau^{+}\tau^{-}$, and $\gamma\gamma$ are combined to produce 95\% CL upper limits on SM Higgs production as a function of mass.   Current Tevatron limits are shown, including a new exclusion for SM Higgs masses between 158 and 175 GeV$/c^2$.  We also present prospects for future sensitivity.
}

\FullConference{35th International Conference of High Energy Physics - ICHEP2010,\\
		July 22-28, 2010\\
		Paris France}

\begin{document}

\section{Introduction}

The discovery of the Higgs boson will resolve the longstanding 
question of how the electroweak symmetry is broken, 
and how fermions and bosons acquire mass.  The standard model 
of particle physics hypothesizes that a non-zero 
scalar field with four degrees of freedom permeates the universe, such that the $W^+$, $W^-$, and $Z$ boson gain mass through three of these 
degrees of freedom, while the fourth degree of freedom becomes a new 
scalar boson.  The BEHHGK (pronounced "beck") mechanism and 
boson, named after the authors (Breit, Englert, Higgs, Hagen, Guralnik, Kibble), 
 was proposed in 1964~\cite{BEHHGK}, and is often referred to as the Higgs mechanism and boson.   Finding the Higgs boson confirms that the Higgs field exists, and is the subject of an intense search at the Tevatron collider.  

Relationships between measurable electroweak parameters within the standard model, in conjunction with direct searches~\cite{Barate:2003sz}, constrain the Higgs boson mass to be between 114 GeV\footnote{c is set to 1 in this proceeding} and 185 GeV at the 95\% CL~\cite{Alcaraz:2009jr}.   With enough data, this entire mass range is accessible to the 1.96 TeV center of mass Tevatron proton-antiproton collider, for either observation or exclusion of the standard model Higgs boson.  The two multipurpose detectors at the Tevatron, CDF and D0, are able to reconstruct all of the final state particles and topologies resulting from SM Higgs boson production and decay.  The Tevatron has delivered 9 fb$^{-1}$ of  luminosity each to CDF and D0.  Data collection efficiencies were 85 - 90 \% for this data, and an integrated luminosity of up to 6.7 fb$^{-1}$ have been analyzed for the Higgs boson searches covered in these proceedings.  The 350 nb$^{-1}$ delivered to the 7 TeV proton-proton LHC is not sufficient for Higgs searches, meaning the search for the Higgs boson is currently unique to the Tevatron. 

The CDF and D0 experiments at the Tevatron have been producing combined Higgs searches since 2006.  Each iteration, efforts are made to improve signal acceptance, such as by loosening lepton and $b$ hadron identification requirements, adding backup triggers which select events online using different aspects of the signal signature, and relaxing kinematic selection.    As the signal acceptance is improved, the backgrounds increase and become more difficult to model, and are separated into categories with similar S $ / \sqrt{B}$, such that high S $ / \sqrt{B}$ categories provide the most signal sensitivity, whereas low S $ / \sqrt{B}$ categories provide the best background constraints.  The primary Higgs channels of $H \to W^+ W^-$, $WH \to l \nu b \bar{b}$, $ZH \to \nu \nu b \bar{b}$, and $ZH \to  l^+l^- b \bar{b}$ from CDF and D0 correspond to about 500 Higgs bosons produced at each mass, 114 $ < m_H <$ 185 GeV.

\section{Low mass SM Higgs searches}

\begin{figure}[htbp]
\includegraphics[width=0.3\textwidth]{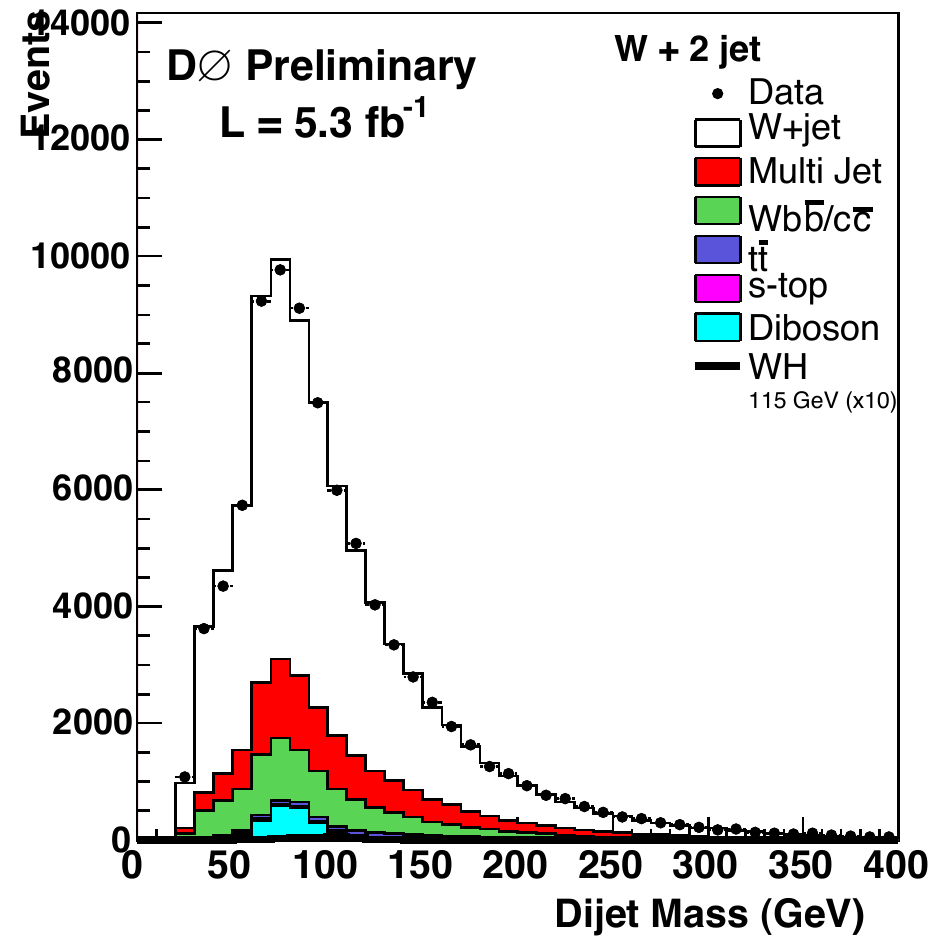} 
\includegraphics[width=0.3\textwidth]{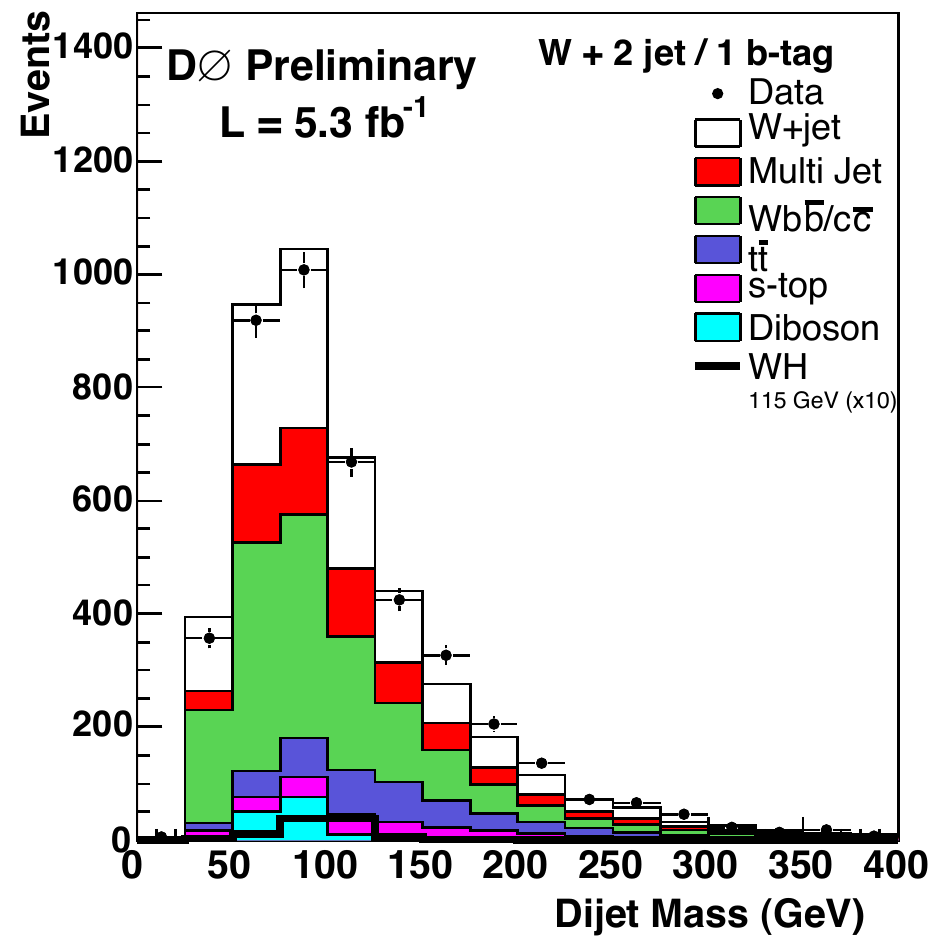} 
\includegraphics[width=0.3\textwidth]{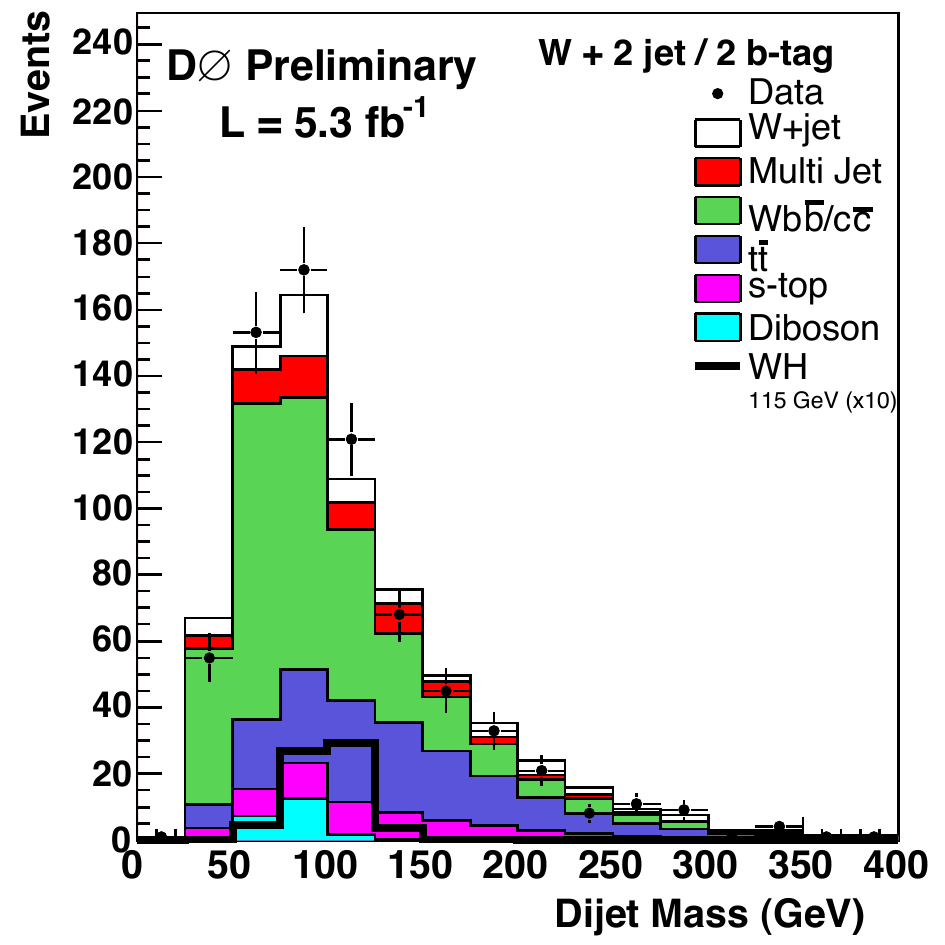}
\caption{\label{fig:mjj}
The dijet mass for $WH \to \ell \nu b \bar{b}$ candidates with no $b$-tags, one $b$-tag, and two $b$-tags (left to right),  demonstrating how the $W$+jets background is reduced while $WH$ signal,  $W+b\bar{b}/c\bar{c}$, and $t\bar{t}$ heavy flavor  backgrounds become more prominent. 
}
\end{figure}

Searches for a "low mass" (m$_H <$ 135 GeV) standard model Higgs boson are presented elsewhere in these conference proceedings for $H \to b \bar{b}$~\cite{bb},  $H \to \tau^+ \tau^-$~\cite{tev-tau},  and $H \to \gamma \gamma$~\cite{tev-gam}.   Here we report on the features of the most sensitive low mass Higgs search channels.   A low mass Higgs boson preferentially decays $H \to b \bar{b}$.  It is most easily identified in events produced via associated Higgs production, $WH$ and $ZH$, when the $W$ and $Z$ decay leptonically, into final states of $WH \to l \nu b \bar{b}$, $ZH \to \nu \nu b \bar{b}$, and $ZH \to  l^+l^- b \bar{b}$ where $\ell = e$ or $\mu$.   In events with a reconstructed $W$ or $Z$ boson and two or more additional jets,  the di-jet invariant mass is used to search for a resonance originating from $H \to b\bar{b}$.   To reduce the background from $W$ and $Z$ production in association with jets, $b$-jets are "tagged" by identifying a secondary vertex or tracks with high impact parameter from $B$ hadron decay.  The dijet mass is shown before and after one and two $b$-tags for $WH \to \ell \nu b \bar{b}$ candidates in Figure~\ref{fig:mjj}.   Expected sensitivity is improved by categorizing events according to the number of $b$-tags and the purity of the applied $b$-tagging algorithms. 

\section{High mass SM Higgs searches}

\begin{figure}[htbp]
\includegraphics[width=0.6\textwidth]{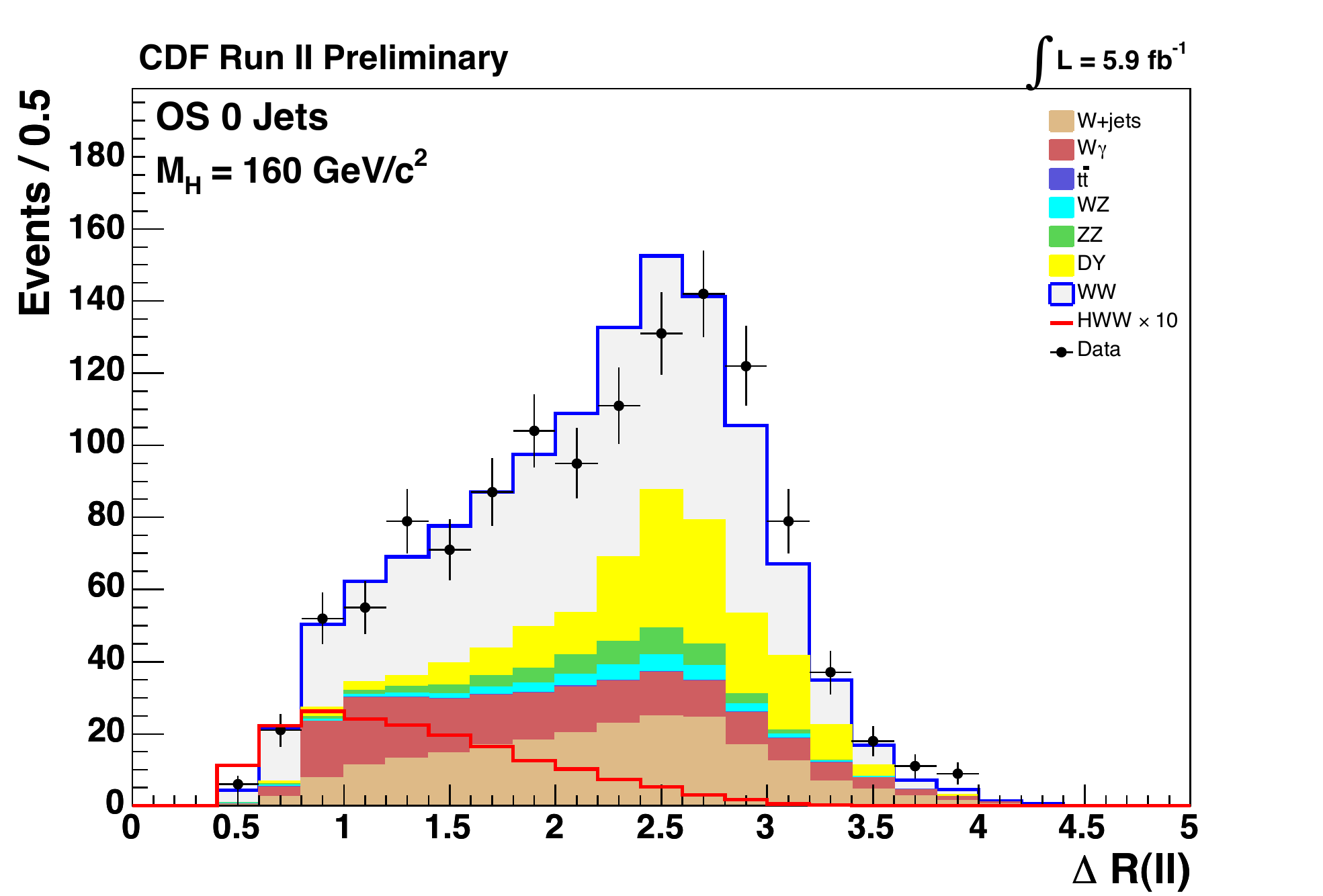} 
\includegraphics[width=0.4\textwidth]{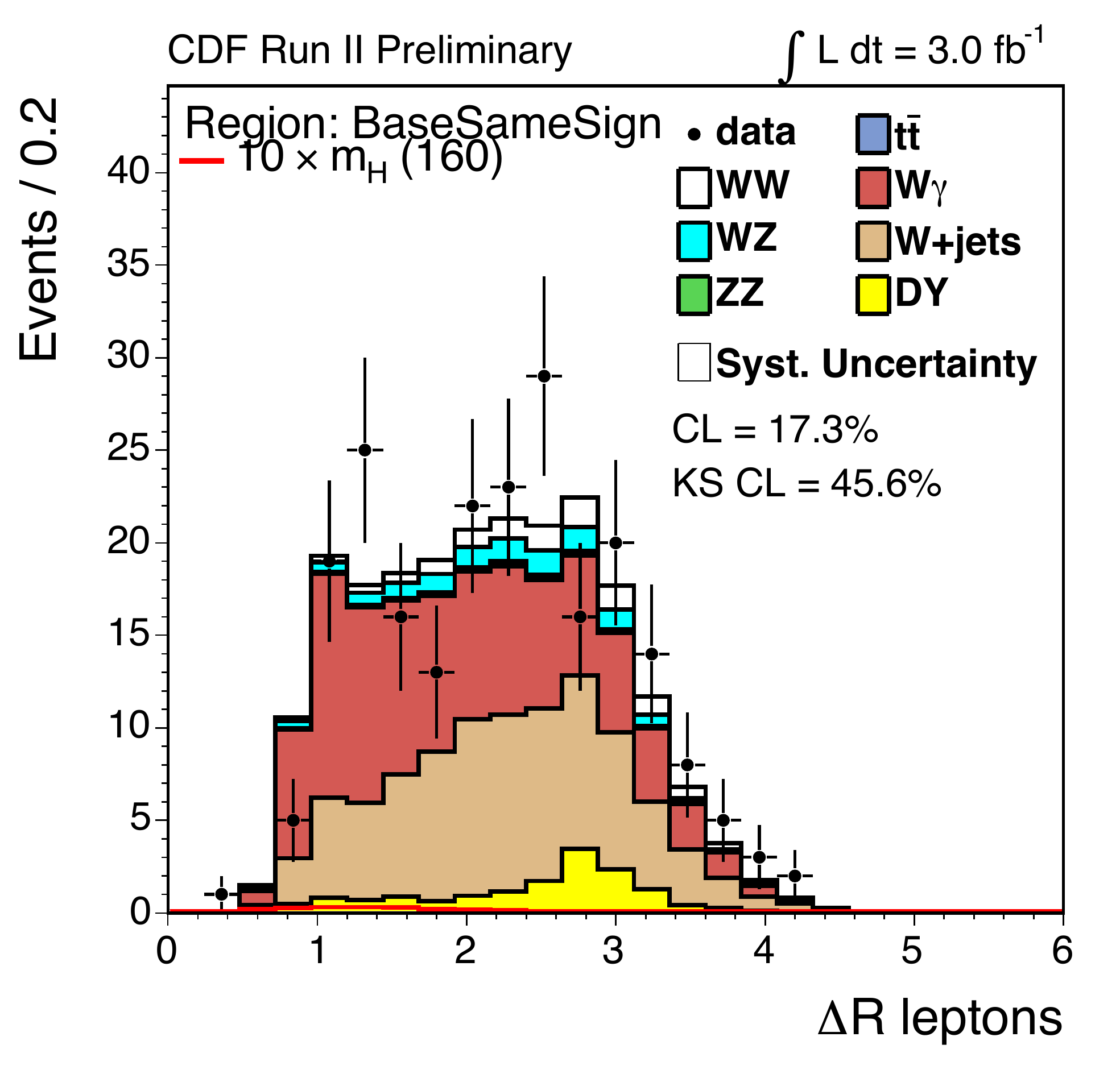}
\caption{\label{fig:HWW-dR}
The $\Delta$R for opposite sign (OS) leptons with no accompanying jets (left) and for the same-sign-lepton control region (right). 
}
\end{figure}

Searches for a "high mass" Higgs boson ($m_H >$ 135 GeV) are reported in detail elsewhere in these proceedings for CDF and D0~\cite{highmass}.  The most sensitive mode at the Tevatron is $gg \to H \to WW \to \ell \nu \ell \nu$ due to the high cross section and well identified final state.  The high mass analysis benefits by separating events into categories according to the number of jets and the number of leptons because of the difference in the kinematics of the signal production mechanisms and primary background processes.  The signal cross section and uncertainties for the dominant signal process $gg \to H$ use state of the art NNLL and NNLO calculations as explained in Ref.~\cite{2010ar}.  Since there are neutrinos in the final state, the invariant mass of the Higgs boson cannot be reconstructed.  The best variable for distinguishing $H \to WW$ from the background is the separation between the charged leptons, dR $= \sqrt{d\eta^2 + d\phi^2}$, which is different for spin-zero $H \to WW$ decay than for spin-one $Z \to WW$ decay as demonstrated in Figure~\ref{fig:HWW-dR}.  Also shown is the equivalent plot for same sign leptons, one of the control regions used to test the background modeling. 

\section{Validation of multivariate analysis techniques}

\begin{figure}[htbp]
\includegraphics[width=0.5\textwidth]{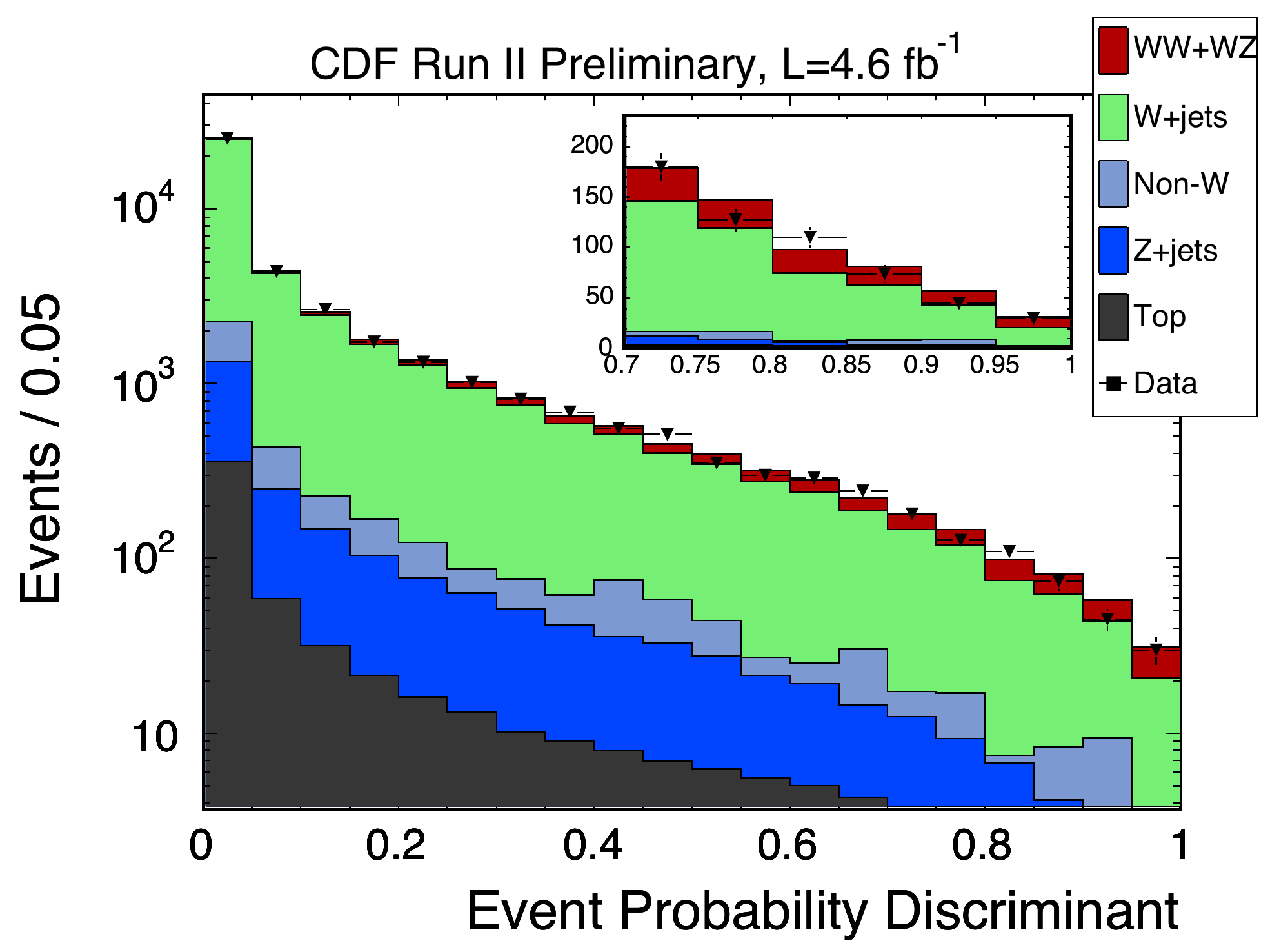}
\includegraphics[width=0.5\textwidth]{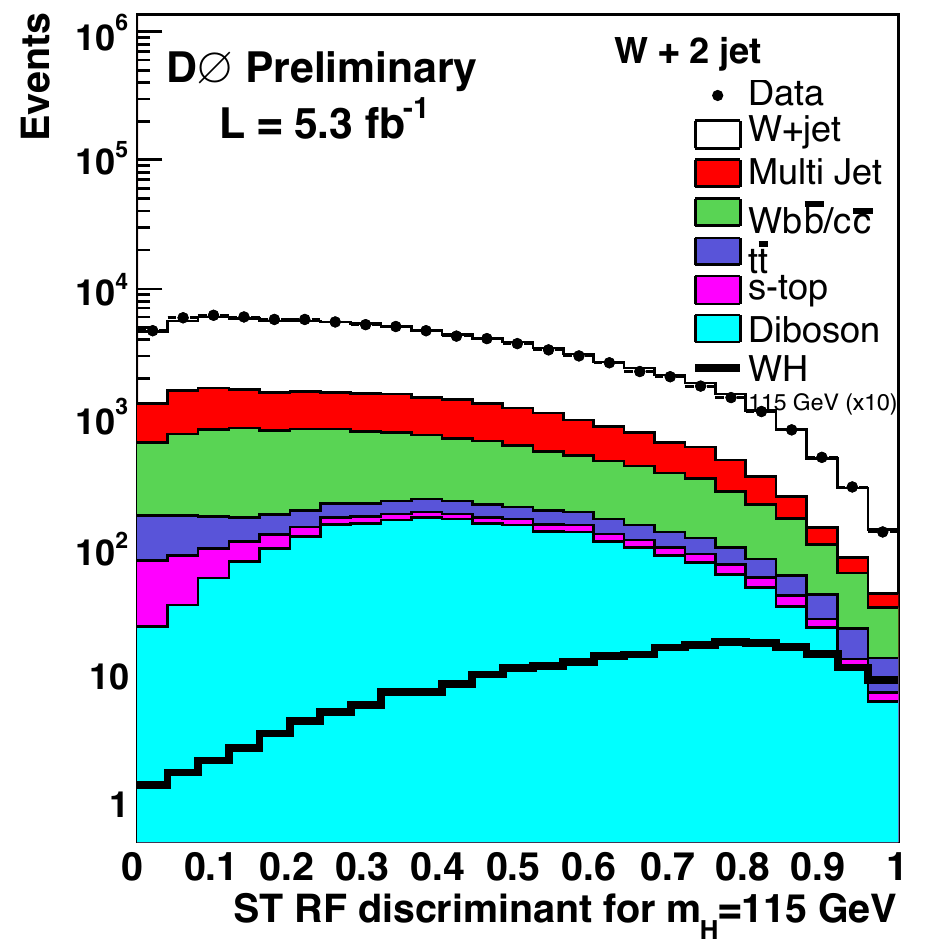} 
\caption{\label{fig:multivariate}
A matrix element event probability discriminant from a diboson search for $WW/WZ \to \ell\nu jj$ (left) and a random forest (RF) of boosted decision trees tested in the pre b-tagged WH control region (right). 
}
\end{figure}

Multivariate techniques are used in all the Tevatron SM Higgs analyses.  The main algorithms used are Neural Networks, Matrix Element probabilities, and Boosted Decision Trees.  By taking advantage of the different correlations for signal and background among the multiple input variables, these techniques typically improve analyses by about 20\% with respect to simply fitting the leading two kinematically distinct variables.  To help validate these multivariate techniques we use them to measure higher statistics standard model processes.   Shown in Figure~\ref{fig:multivariate} are two measurements using multivariate techniques.  We measure $\sigma_{WW+WZ} =$ 16.6$^{+3.5}_{-3.0}$ pb in a $W$+2-jet final state compared to the SM expected cross section of 15.1 $\pm$ 0.8 pb~\cite{Aaltonen:2009vh}, and also 
verify the $W$+jets background process rate and shape by analyzing a $WH \to \ell \nu b\bar{b}$ sample of events before the $b$-tagging requirements. 

\section{Combined SM Higgs search}

\begin{figure*}
  \resizebox{0.55\textwidth}{!}{\includegraphics{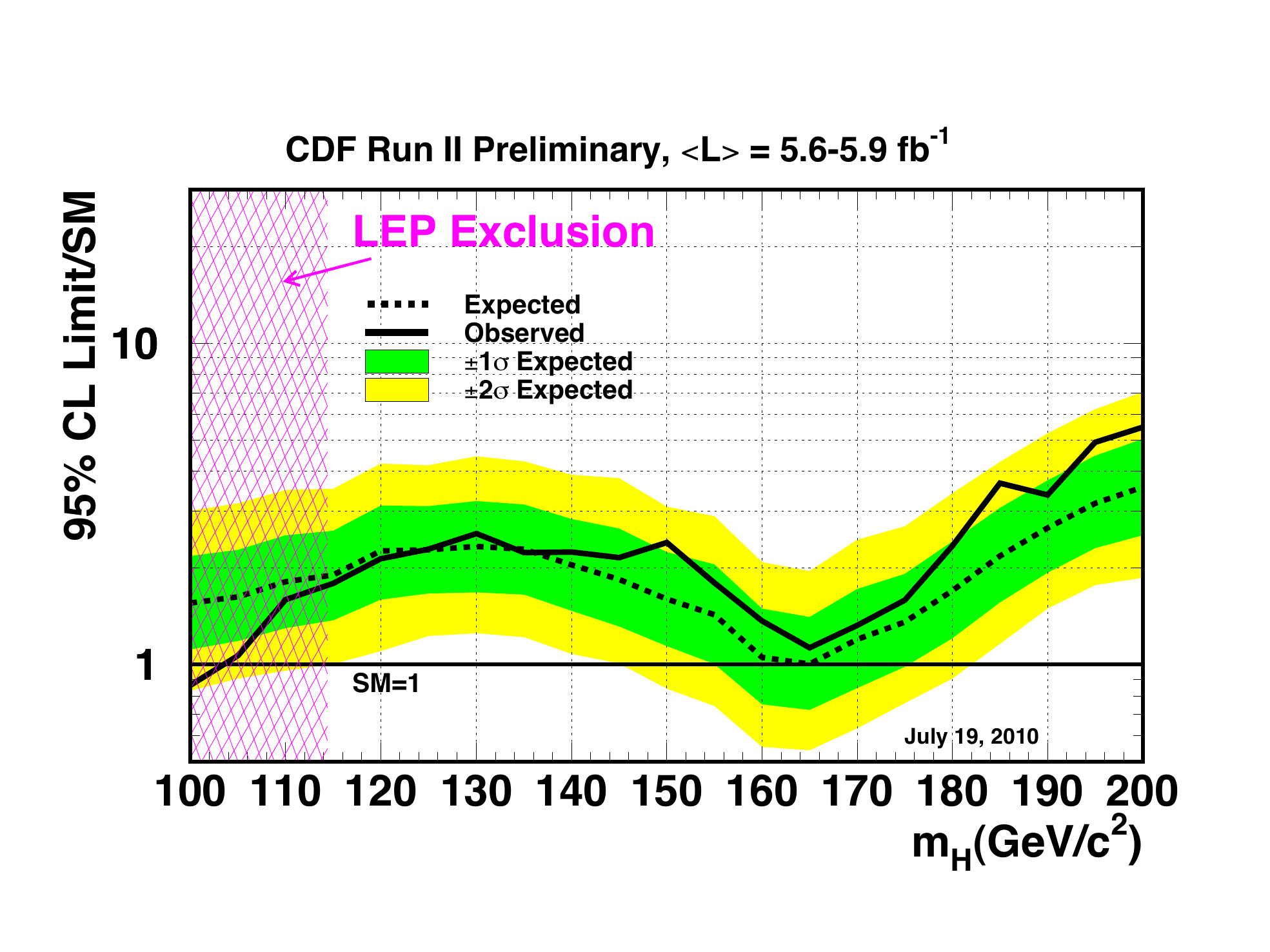}}
  \resizebox{0.45\textwidth}{!}{\includegraphics{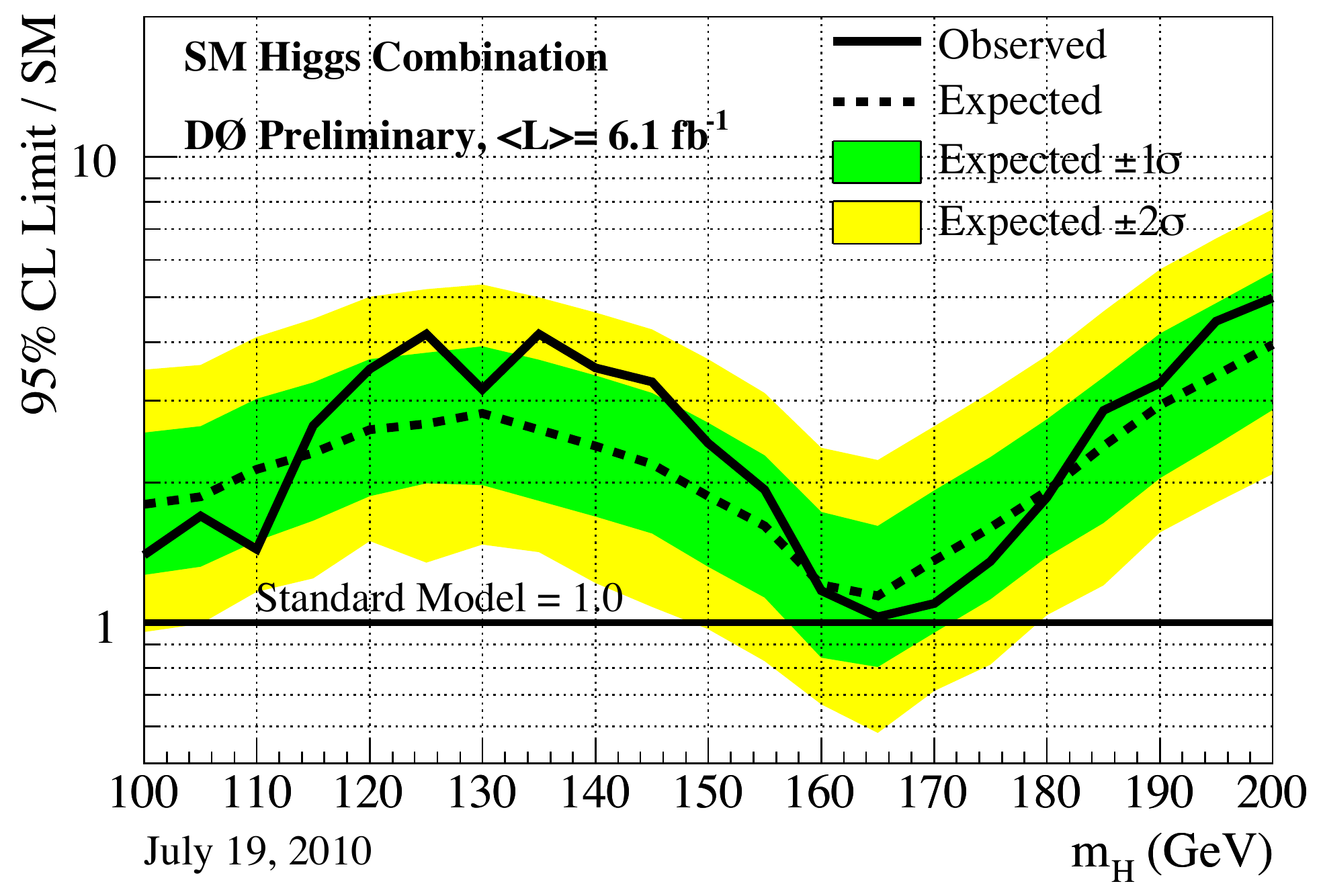}} \\ 
  \resizebox{\textwidth}{!}{\includegraphics{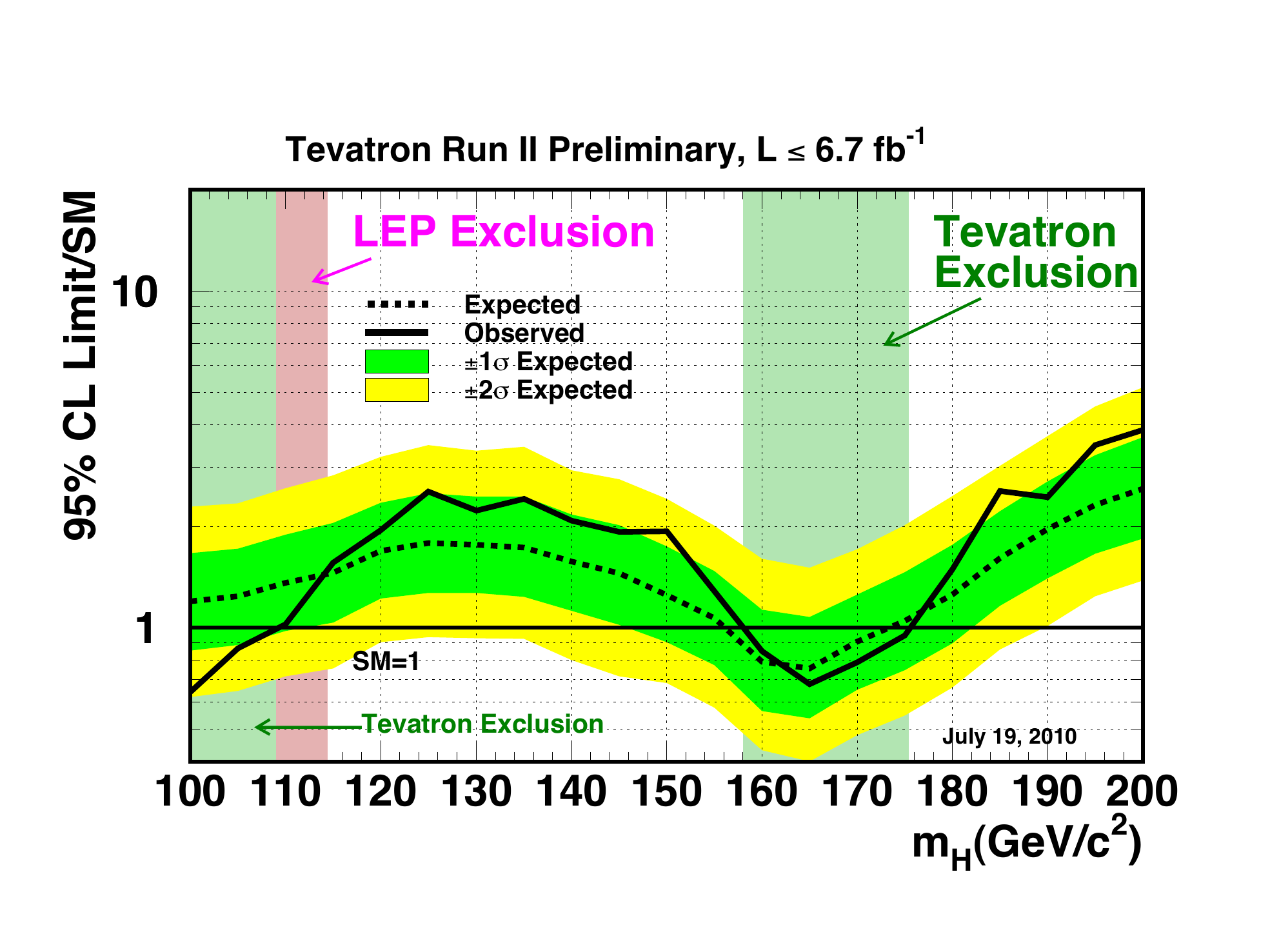}}
\caption{SM Higgs combinations for CDF (left), D0 (right), and the Tevatron (bottom).}
\label{fig:cdf-d0-combo}
\end{figure*}

The CDF combined search and D0 combined search for the SM Higgs boson was presented elsewhere in these proceedings \cite{cdf-combo, d0-combo} and the final results are shown in Figure~\ref{fig:cdf-d0-combo}.   The  95\% CL upper limit on the cross section divided by the SM value of the cross section is shown on the y-axis and is equal to one when standard model sensitivity is achieved.   The dotted line is the expected upper limit on cross section in pseudo-experiments with no signal events, allowing systematic uncertainties to be fit within each pseudo-experiment, and fitting for the maximum signal that can be accommodated at the 95\% CL.  The green and yellow bands represent the one and two $\sigma$ variations of the expected limits.  The solid line is the upper limit observed in actual data.  CDF achieves expected sensitivity to the Higgs boson for m$_H =$ 165 GeV, while D0 almost achieves observed sensitivity for  m$_H =$ 165 GeV.  CDF is able to exclude 100 < m$_H$ < 102 GeV. 

CDF and D0 perform a joint search for the Higgs boson where shared systematic uncertainties are kept correlated (bottom of Fig.~\ref{fig:cdf-d0-combo}).  The observed exclusion is 158 $< m_H <$ 175 GeV which is consistent with the expected exclusion of 156 $< m_H <$ 175 GeV.   The combined search is also able to exclude Higgs masses below 109 GeV.   At 115 GeV, just above the LEP limit, the expected (observed) exclusion is 1.45*SM (1.56*SM).    For $m_H =$ 165 GeV, where the high mass exclusion is strongest, $H \to W^+W^-$ dominates, but for $m_H =$ 115 GeV, the sensitivity comes from a combination of analyses (see Table~\ref{tab:limit}).  

\begin{figure*}
  \resizebox{0.5\textwidth}{!}{\includegraphics{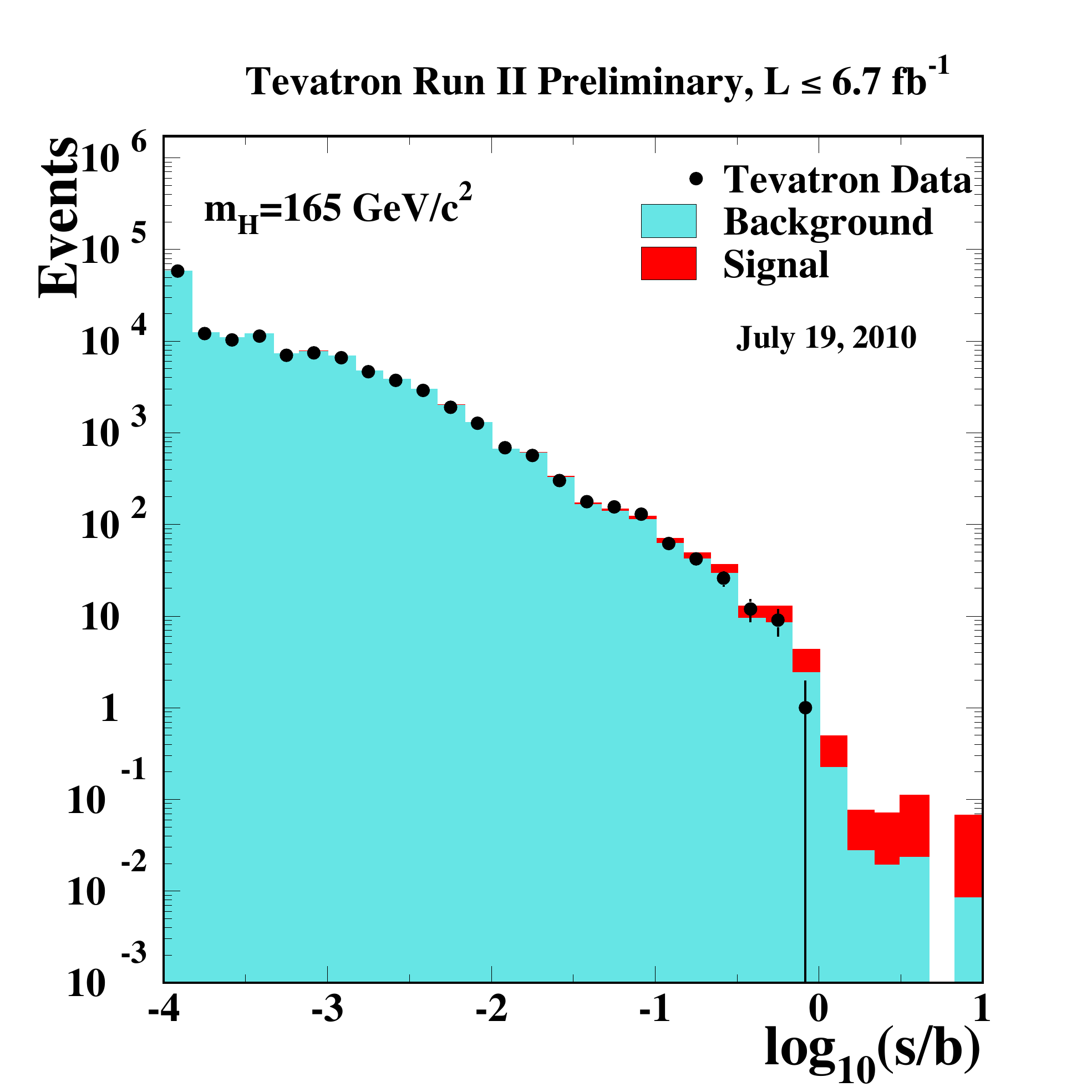}}
  \resizebox{0.5\textwidth}{!}{\includegraphics{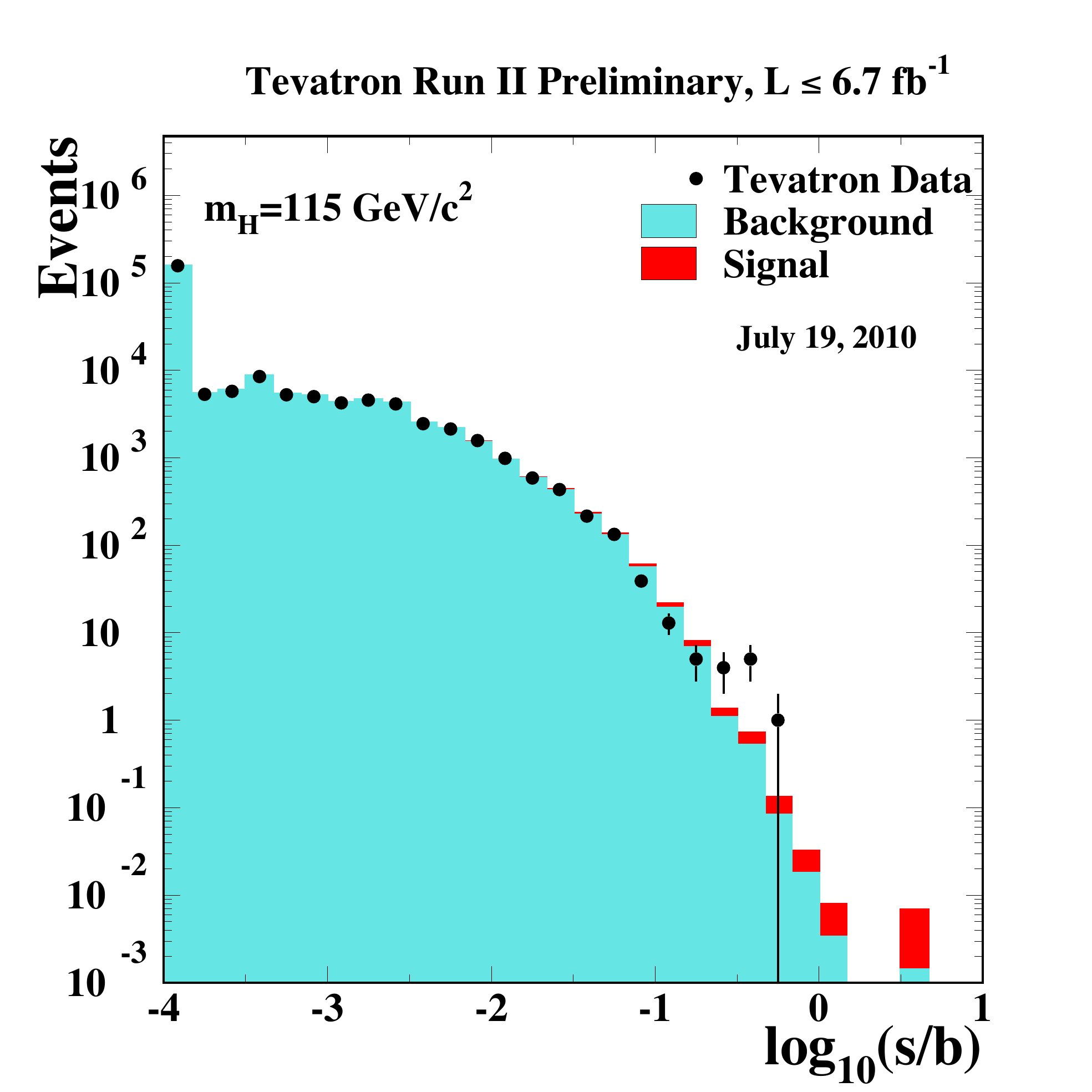}}
\caption{ On left, for m$_H =$ 165 GeV, the data from all search channels sorted into bins of varying $S/B$.   On right, same for m$_H =$ 115 GeV .   }
\label{fig:logSB}
\end{figure*}

We can collectively view the dozens of discriminant outputs from the multiple search channels by gathering all histogram bins from all sub-channels, summing up those bins with the same $S/B$ into the same bins, and then sorting the bins from lowest to highest $S/B$.  The distribution is shown for m$_H =$ 165 GeV and m$_H =$ 115 GeV in Figure~\ref{fig:logSB}.  For 165 GeV, the data clearly prefer the background-only model, while for m$_H =$ 115 GeV an excess of events in bins with very high $S/B$ and a deficit of events with slightly lower $S/B$ result in the observed limit being approximately equal to the expected limit. 

%
%


\begin{table}
\caption{Tevatron analyses ordered by the expected upper limit at 95\% CL for 
a Higgs boson search mass of 115 GeV.  }
\label{tab:limit} 
\begin{center}
  \begin{tabular}{cccc}
   \hline
   \hline
   Analysis     &  Experiment & Expected &  Integrated            \\
      channel    &  &  limit  $\sigma$/$\sigma_{SM}$ @ 115 GeV & Luminosity (fb$^{-1}$)  \\
   \hline                                      
   $WH \to \ell \nu b\bar{b}$    & CDF   & 3.4    &      5.7                           \\
   $ZH/WH \to \met b\bar{b}$    & CDF   & 4.0    &     5.7                               \\
    $ZH/WH \to \met b\bar{b}$     & D0   & 4.2    &      6.4                              \\
   $WH \to \ell \nu b\bar{b}$    & D0   & 4.8    &      5.3                               \\
   $ZH \to \ell \ell b\bar{b}$    & CDF   & 5.5    &      5.7                               \\
    $ZH \to \ell \ell b\bar{b}$    & D0   & 5.7    &     6.2                               \\
    $H \to WW$ & CDF & 10.6 & 5.9 \\
    $H \to WW$ & D0 & 12 & 6.7 \\
    $H \to \tau\tau$ & D0 & 16 & 4.9 \\
    $ZH/WH \to qq b\bar{b}$    & CDF   & 18    &     4                              \\
     $H \to \gamma \gamma$ & D0 & 18.5 & 4.9 \\
     $H \to \gamma \gamma$ & CDF & 21 & 5.4 \\
     $H \to \tau\tau$ & CDF & 25 & 2.3 \\
    $ttH$ & D0& 45 & 2.1 \\
	\hline
 Total & CDF + D0 & 1.45 &  5.8 \\
    \hline
    \hline
   \end{tabular}
 \end{center}
\end{table}

\section{Beyond SM Higgs searches}

\begin{figure*}
  \resizebox{0.8\textwidth}{!}{\includegraphics{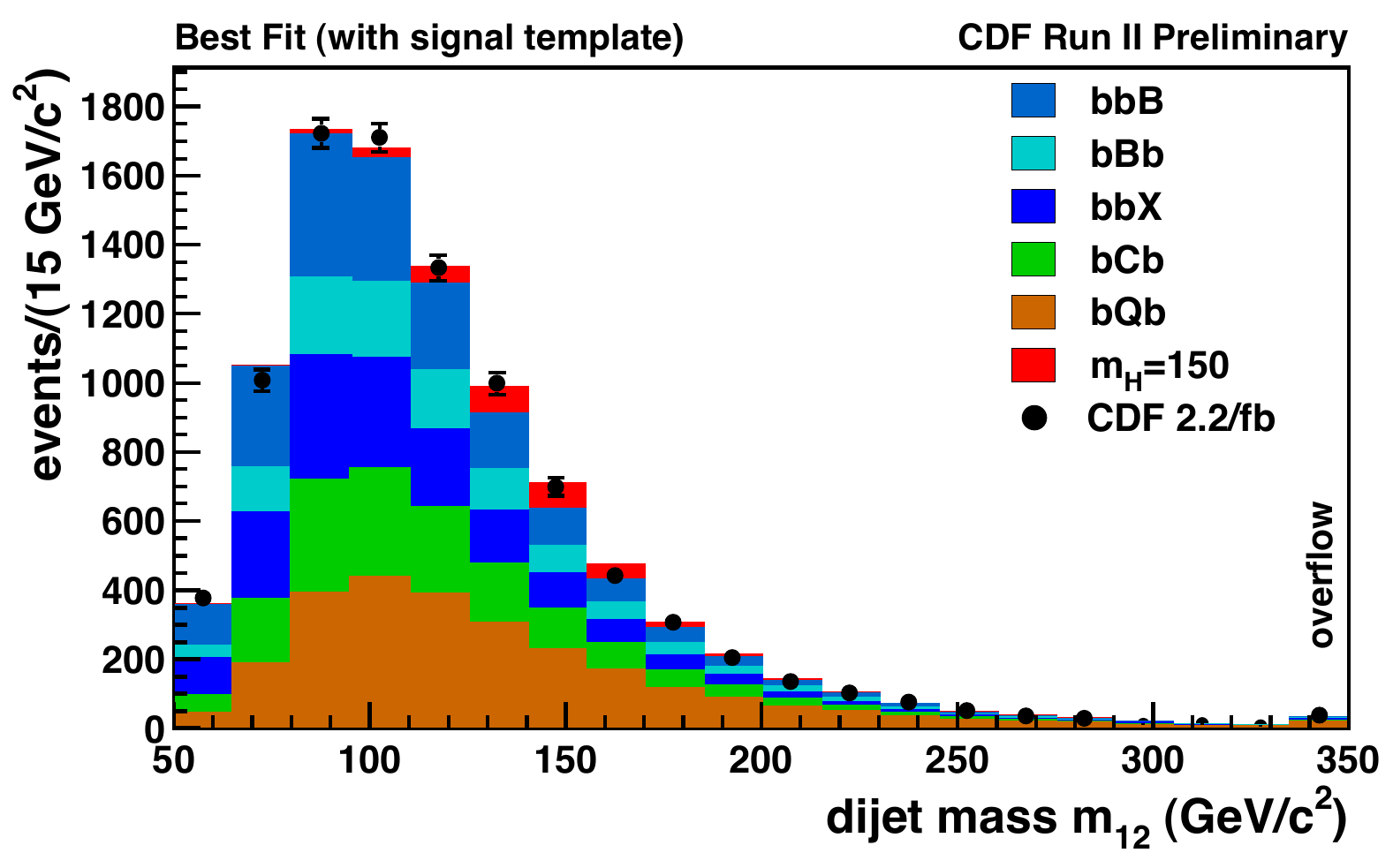}}
\caption{ Search for the MSSM in $H+b \to bb+b$ search channel at CDF shown with best fit to signal plus background hypothesis.}
\label{fig:mssm-3b}
\end{figure*}

Beyond standard model Higgs bosons searches are presented in detail elsewhere in these conference proceedings~\cite{tev-bsm}.  The MSSM predicts enhanced Higgs production cross sections and decays to down-type fermions for high values of tan $\beta$, the ratio of the vacuum expectation value for up-type and down-type fermions, thereby enhanacing rates for $H \to b\bar{b}$ and $H \to \tau\tau$ final states.  The Tevatron has searched for such signatures primarily using the invariant mass of the $b$-jets or the invariant mass of the visible $\tau$ decay products as discriminants and set limits close to the theoretically interesting value of tan $\beta =$ 30, which is approximately the ratio of m$_{top}$ to m$_{b}$.  A small excess of 2 $\sigma$ (including a trials factor) is seen in the CDF $bb$ search (Fig.~\ref{fig:mssm-3b}).

\section{Outlook}

\begin{figure*}
  \resizebox{0.8\textwidth}{!}{\includegraphics{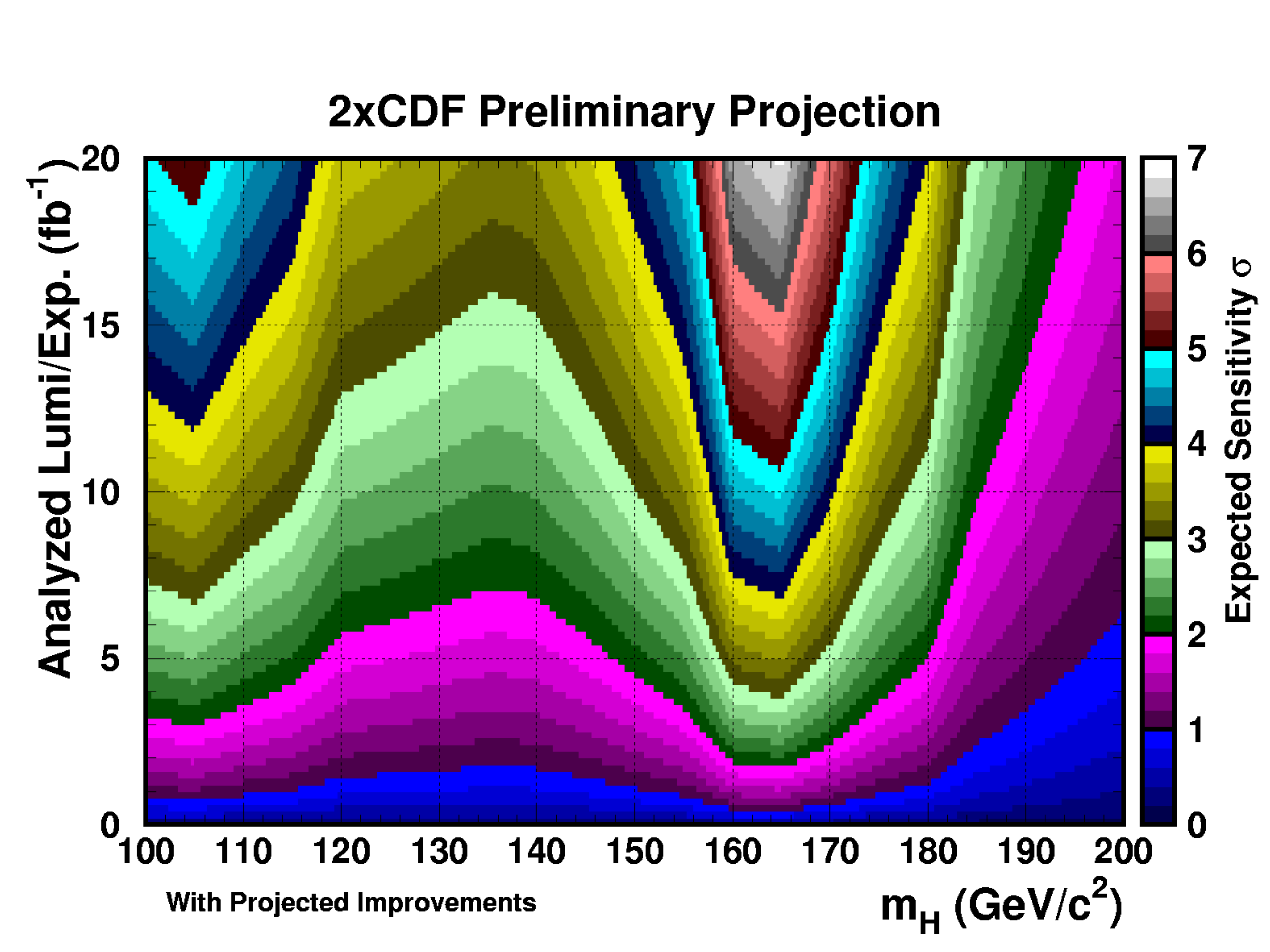}}
\caption{ Projections of expected Higgs sensitivity, assuming a factor of 1.5 improvement in the expected limits at each mass point, and assuming D0 and CDF have equal sensitivity and equal analyzed integrated luminosity.  Note that much of this factor of 1.5 has already been achieved at high mass.  }
\label{fig:proj2xcdf}
\end{figure*}

Figure~\ref{fig:proj2xcdf} shows the expected a priori signal sensitivity as a function of mass and analyzed integrated luminosity per experiment.  This includes a factor of 1.5 improvement in the expected limit based on a range of improvements from lepton identification efficiency, b-tagging efficiency, triggering, and jet energy resolution.  The full dataset for Run II of the Tevatron will be collected by September 2011, and is expected to be about 10 fb$^{-1}$ of analyzed luminosity per experiment.   The projected sensitivity with this dataset is 3 $\sigma$ for m$_H =$ 115 GeV. The 
sensitivity would be expected to be at least 2.4 $\sigma$ across the full Higgs boson mass range of 100 to 185 GeV.  The possibility of running the Tevatron for another 3 years would allow for about 17 fb$^{-1}$ of data to be analyzed per experiment.   This would increase the expected sensitivity to 3 $\sigma$ across the full mass range, 4 $\sigma$ at 115 GeV. 

%

\section{Summary}

The CDF and D0 experiments at the Tevatron have a comprehensive and aggressive search program for the Higgs boson which can be divided into three main categories : high mass, low mass, and beyond standard model searches.   The newest high mass combination from the Tevatron excludes 158 $< m_H <$ 175 GeV using a dataset of up to 6.7 fb$^{-1}$.  The low mass primary channels achieved 1.45*SM (1.56*SM) expected (observed) exclusion for $m_H =$ 115 GeV and a 95\% CL exclusion for masses between 100 GeV and 109 GeV.  Tevatron MSSM Higgs searches are setting limits at theoretically motivated values of tan $\beta$, and a  $H \to b \bar{b} + b$ search with 2.2 fb${-1}$ has a 2 $\sigma$ excess at m$_H =$ 140 GeV.  The Tevatron is scheduled to deliver up to 12 fb$^{-1}$ by its conclusion in September 2010, providing about 10 fb${-1}$ of analyzable data for Higgs analysis.   With this dataset, a full combination of Higgs searches will have  at least 2.4 $\sigma$ level sensitivity for 100 $< m_H <$ 185 GeV.  This would increase to 3 $\sigma$ were the Tevatron to run an additional three additional years as is the Tevatron Run III proposal.

\end{document}